\documentstyle[12pt]{article}
\begin{document}
\begin{center}
\vskip 2.0 truecm
{\bf \Large   Phenomenological parametrization of the charged particle
            multiplicity distributions in restricted rapidity
 intervals in $e^+e^-$ annihilation into hadrons
 and $e^+p$ scattering at HERA\footnote{This 
            work was  supported in part 
            by INTAS, contract INTAS-93-3602 } }
\vskip 1.5 truecm
 {\sc   O. G. Tchikilev\footnote{ E-mail: tchikilov@mx.ihep.su} }
\vskip 0.4cm
  {\em Institute for High Energy Physics  \\
   142284, Protvino, Russia\/}
\end{center}
\vskip 1.5 truecm
\begin{abstract}
\noindent
It is shown that  charged particle multiplicity distributions
in restricted (pseudo)rapidity intervals in $e^+e^-$ annihilation
and in $e^+p$ scattering at HERA are quite well described by the
modified negative binomial distribution and
its  simple extension. 
\end{abstract}
\vskip 1.5cm
(  to be published in Phys. Lett. B)
\newpage
\pagestyle{plain}
\subsection*{ }
  Recently it has been shown~[1-5]  that
negatively charged particle multiplicity
 distributions in $e^+e^-$ annihilation are well described by the 
 modified negative binomial distribution (MNBD) with the parameter
 $ \Delta $ being practically energy independent and the parameter 
 $k$ approaching the value 7. It has been demonstrated  also
  that the MNBD fairly well
 describes multiplicity 
 distributions in lepton-nucleon interactions with a similar behaviour of the
 parameters~[6]. The aim of this paper is to show that the MNBD and its
 simple extension 
 describe quite well
 the charged particle multiplicity distributions in restricted
 rapidity intervals in $e^+e^-$ annihilation into hadrons~[7-12], and
  charged particle multiplicity distributions in restricted pseudorapidity
 intervals in $e^+p$ scattering at HERA energies~[13].

  The MNBD is defined by the probability generating function
\begin{equation} 
           M(x) = \sum_{n}{P_{n}x^{n}} = 
   {\biggl({{1+\Delta\, (1-x)}\over{1+r\,(1-x)}}\biggr)}^{k},
\end{equation}
where  $P_n$ is the probability to produce ~$n$~ particles. 
 The MNBD  can be considered as a convolution of the Newton
 binomial (the numerator in (1)) and the negative binomial (the denominator
in (1)). The probabilities
 $P_{n}$ for the MNBD can be computed also using iteration relations given
 in~[4] or the formulae given in~[14]. We have checked
 that all three methods give the same $P_{n}$ values
 within the computer precision
 limits.

 It has been assumed in~[1,3,4]  that the parameter
 $k$ is the number of
 sources of particle production at some initial stage of the interaction; these
 sources develop independently of each other according to some branching
 process, producing intermediate neutral clusters. In this model the
 parameter $-\Delta$ is equal to the cluster decay probability into a
 charged hadron pair ($1+\Delta$ is the probability of cluster decay into
 pair of neutral hadrons). One can associate these sources with the widely
 discussed clans~[15] or superclusters~[16].
 A priori it is not evident that the same
 distribution should be valid for restricted rapidity intervals; indeed
 the initial sources can move with respect to each other and therefore
 contribute differently to the given rapidity interval, the same 
 can happen for the intermediate neutral clusters. Let us assume  that 
 a fixed number of initial sources (less than or equal to $k$) contributes
 to a given rapidity interval $(y_1,y_2)$, and both charged decay products 
 of each neutral cluster hit  the rapidity interval with the
 probability $\epsilon_2$ and only one charged particle hits the interval
 with the probability $\epsilon_1$. The probabilities $\epsilon_1$ and
 $\epsilon_2$ are assumed to be the same for all neutral clusters,
 whatever their source.
In this case the probability generating
 function for the charged particle
 multiplicity distribution has the form (1), where $x$ is replaced by
\begin{equation} 
    \varphi_{2}(x) = 1 -\epsilon_{1}(1-x)-\epsilon_{2}(1-x^2).
\end{equation}
 The probability $\epsilon_{1}$ should increase with the size of the
 rapidity interval up to a maximum, and then go down to  zero; whereas
 the probability $\epsilon_2$ should increase monotonically from zero to
 one. One expects that this rapidity dependence
 can  naturally explain odd-even
 fluctuations for rapidity intervals covering nearly
  the whole phase space. One
 expects also that the parameter $k$ increases from some initial value
 (may be 1) to the maximum, observed over the whole phase space. For some
 rapidity intervals, when contributions from configurations with 
 different number of sources are equally important, the above assumptions
 seem to be crude; in this case one can use a sum of distributions with
 different $k$ values\footnote{This approach, for example, is used in~[17],
  where multiplicity distributions are parametrized by a
 weighted superposition  of two negative binomials.}.
 
 Analytical formulae for $P_n$ can be obtained  using the
 decomposition of the quadratic term in the the denominator
 of the probability generating function into the product of two linear
 terms. For $k=1$ it leads to the following formulae for the probabilities:
\begin{equation}
   P_{0} = \frac{1 + \Delta\, (\epsilon_1 + \epsilon_2 )}
                {1 + r\, (\epsilon_1 + \epsilon_2 )}\, ,
\end{equation}
\begin{equation}
  P_{i} =\frac{(1-\Delta/r)\, (\beta_{1}^{(i+1)} - \beta_{2}^{(i+1)})}
         {(1+r\, (\epsilon_1 + \epsilon_2 ))\, (\beta_1 -\beta_2)}\, ,~ ~i>0,
\end{equation}
 where
\begin{equation}
 \beta_1 = \frac{b_1}{2} + \sqrt { \frac{b_{1}^{2}}{4} + b_2}\, ~,
 ~\beta_2 = \frac{b_1}{2} - \sqrt { \frac{b_{1}^{2}}{4} + b_2}
\end{equation}
 and
\begin{equation}
 b_1 = \frac{r \epsilon_1}{1+ r\, (\epsilon_1 + \epsilon_2)}\, ~,
 ~b_2 = \frac{r \epsilon_2}{1+ r\, (\epsilon_1 + \epsilon_2)}\, .
\end{equation}
 For $k$ greater than one we have calculated the $P_n$ as a $k$-th
 convolution of the multiplicity distribution with $k=1$, given by the
 formulae (3) and (4). It has been checked that for $\epsilon_2 =0$
 these calculations give the same results as the MNBD formulae.

 Probabilities $P_n$ can be computed also using the recurrence relations
 for Gegenbauer polynomials $C_{n}^{\lambda}(t)$ with the generating
 function
\begin{equation}
  \sum_{n=0}^{\infty} C^{\lambda}_{n}(t) z^{n} = (1 - tz + z^2 )^{-\lambda}
\end{equation}
 (see for example~[18]). The case  $\Delta=0$ gives
\begin{equation}
  P_{n+1} = \frac{n+k}{n+1}b_1 \, P_n + \frac{n+2k-1}{n+1}b_2 \, P_{n-1}
\end{equation}
 with $b_1$ and $b_2$ given by (6) and with initial conditions
\begin{equation}
 P_0 = (1+r(\epsilon_1 +\epsilon_2))^{-k} \, ~,~ P_1 = k b_1 P_0 \, .
\end{equation}
 These recurrence relations can be used also
 for the case with $r=0$\, if one replaces
 $k$ by $-k$ and $r$ by $\Delta$, and the final probabilities can be calculated
 by the convolution of these two distributions with $\Delta=0$ and $r=0$.
 We have checked that this  method gives the same results as the method
 based on  formula (4).
 
  The experimental data have been parametrized by the MNBD ($\epsilon_2=0$)
 and its extension with $\epsilon_2 \neq 0$.
 The parameter $\Delta$ was fixed at the value $\Delta=-0.76$ taken from
 fits of the $e^+e^-$ data~[4,5] and the
 parameter $r$ was calculated from the mean charged multiplicity $<n>$
using the relation
\begin{equation}
 r = \Delta + \frac{<n>}{k\, (\epsilon_1 + 2 \epsilon_2)}\, .
\end{equation}
  The integer parameter $k$
 has been tested in the interval from 1 to 9, and the parameters $\epsilon_1$
 and $\epsilon_2$
 have been assumed to be nonnegative. One should note that the parameter
 $\Delta$
  cannot be obtained directly from fits, indeed the same $P_{n}$
 can be obtained for other $\Delta$
 values by simple rescaling of the $\epsilon_1$,
  $\epsilon_2$ and $r$.  

  Tables 1 and 2 give the results of the  fits to the 
 charged particle multiplicity distributions in the symmetric rapidity
 intervals $|y|<y_{cut}$ for $e^+e^-$ annihilation
  for the TASSO~[8], DELPHI~[9] and ALEPH~[11] data\footnote{
 The $P_n$ for the TASSO measurements~[8] were calculated from the
 KNO scaling function $\psi (z)$, compiled in the
 COMPAS data base, IHEP, Protvino.}.
%
%
 In some cases when the results with $\epsilon_2 \neq 0$ are not given, the
 fits have stopped on the boundary $\epsilon_2=0$.

   One can see from tables 1 and 2 that the quality of the fits 
is quite qood for
all data excluding the DELPHI measurement~[9].
 It is necessary to  note that the $\chi^2/NDF$ values for the fits
 should be considered just indicative, since the full covariance matrix
 is not published and therefore the proper treatment of the correlations
 between measurements of the neighbour multiplicities is not 
 possible\footnote{Typical uncertainties~(r.m.s.) in measured multiplicities
 in recent high energy experiments range from $\sim 2$ at $n_{ch}=8$ to
 $\sim 4$ at $n_{ch}=30$~[11].}.
  Nevertheless, for the TASSO measurements one can compare these values with
 the $\chi^2/NDF$ values obtained for 
 negative binomial fits~[8] made to the unfolded distributions. 
 The $\chi^2/NDF$ ratios for MNBD fits in general are
  smaller than for negative binomial fits. Especially it is true for the
 statistically most significant TASSO measurement at $\sqrt s =34.8$~GeV.
  The parameter $k$  tends to rise with  increasing 
 size of the rapidity interval. 
The  parameters $\epsilon_1$ and $\epsilon_2$ in general follow the
 expected dependence on the size of the rapidity interval. 

 The big $\chi^2/NDF$ ratios for the DELPHI measurements~[9] are explained
 by the shoulder structure caused by the superposition of events with 
 different number of jets (for the discussion see [17,19]). This shoulder
 structure, well described by the Monte Carlo program JETSET, is
 parametrized in~[17] by a weighted superposition of two negative
 binomials. Following this approach we have parametrized multiplicity
 distributions in restricted rapidity intervals by the weighted
 superposition of two modified negative binomials. Let us denote the
 probabilities $P_n$ for MNBD with the form (2), and $\epsilon_2=0$ by
 $p_n (k,\Delta,r_1,\epsilon_1)$. Then the weighted sum is given by
\begin{equation}
   P_n =\alpha p_n(k,\Delta,r_1,\epsilon_1) +
        (1-\alpha) p_n(k+1,\Delta,r_2,\epsilon_1) \, ,
\end{equation}
 where $\alpha$ is the probability for the contribution of the $k$
 sources with the parameter $r_1$, whereas $1-\alpha$ is the 
 probability for the contribution
of the $k+1$ sources with the parameter $r_2$.
The results of fitting the multiplicity distributions measured by
 DELPHI~[9] in symmetric rapidity intervals by the weighted sum (11)
are given in  table~3. One
can see the $\chi^2/NDF$ ratios are close to one; nevertheless it is
 necessary to note that the fits have four free parameters. We have
tried also to fit by this parametrization the multiplicity distributions~[9]
in asymmetric rapidity intervals $0<y<y_{cut}$ ( or $-y_{cut}<y<0$). The
results are given in  table~4. The $\chi^2/NDF$ ratios are near 1 for
all intervals, excluding the intervals with $y_{cut}=1$ and $y_{cut}=1.5\,$: for
these intervals, the ratios are near 2.

 The results of the  fits of the MNBD,  and its extension with
 $\epsilon_2 \neq 0$,
to charged particle multiplicity distributions
in the single hemisphere for $e^+e^-$ 
annihilation into hadrons~[7-10,12] are given in  table~5. The 
$\chi^2/NDF$ ratios are of  order 1 for practically  all measurements.
 One should note also  
 that the  fits to the DELPHI measurements~[12] for $b\overline{b}$
 events and all flavours
have smaller $\chi^2/NDF$ ratios than the ``shifted'' negative binomial 
fits given in~[19]\footnote{The shifts in multiplicity 
 are due to the decays of the heavy $b$-quarks.}. 
For example, the MNBD fit for the $b\overline{b}$
sample gives $\chi^2/NDF=13.3/27$, and the  fit with $\epsilon_2 \neq 0$ 
gives $\chi^2/NDF=7.9/26$,
 whereas the best negative binomial fit
with a shift in multiplicity by 2 units has $\chi^2/NDF=20/26$~[19].

It is of interest to note that the $k$ values for single hemispheres
are close to the parameter $k$ for MNBD fits to the
 multiplicity distributions in the whole phase space~[4,5], whereas
 the parameter $\epsilon_1$ is rather close to the value 0.5. The second
 observation can be explained if one assumes that the intermediate neutral
 clusters hit one or other hemisphere with equal probability.

 The results of the  fits to the H1 data~[13] in the pseudorapidity
 intervals $1<\eta<\eta_{cut}$ are given in  table~6. The $\chi^2/NDF$
 values are in general close to 1 both for $\epsilon_2 =0$ and for
 $\epsilon_2 \neq 0$. It is of interest to note the the fits with
 $\epsilon_2 \neq 0$ have smaller $\chi^2/NDF$ than the MNBD fits mainly
 for the energy  $W$ above 150~GeV. 

 In conclusion, it is shown that the MNBD and its simple extension 
  describe  multiplicity data in the restricted (pseudo)rapidity
 intervals for $e^+e^-$ annihilation and $e^+p$ deep inelastic scattering
 at HERA quite well. The DELPHI measurements~[9] in symmetric and
 asymmetric rapidity intervals
 are fairly well parametrized by the weighted superposition of two
 modified negative binomial distributions with the parameters
 $k$ and $k+1$. The dependence of the parameters on the size of the 
 rapidity interval is explained by simple assumptions, extending the
 model proposed in~[1,3,4]. 
 Futher checks of this phenomenological model can be done,
 for example, by  studying
  the multiplicity distributions for particles
 of the same charge.

\subsection*{Acknowledgements}
\vskip 3mm

 I am indebted to A.~De Angelis and W.~Venus 
for reading the manuscript and for the suggested improvements. The use
of the COMPAS data base is acknowledged.

\vskip 1cm
\newpage

%
%
%
\vskip 0.5cm
\newcommand{\qq}{\mbox{$Q^{2}$}}
\newcommand{\en}{\mbox{$\sqrt s$}}
\newcommand{\np}{\mbox{$n_p$}}
\newcommand{\pnch}{\mbox{$P(n_{ch})$}}
\newcommand{\nch}{\mbox{$n_{ch}$}}
\newcommand{\avm}{\mbox{$<n_{-}>$}}
\newcommand{\avn}{\mbox{$<n_{-}>$(fit)}}
\newcommand{\axi}{\mbox{$\chi^2$/NDF}}
\newcommand{\dda}{\mbox{$\times 10^{-1}$}}
\newcommand{\ddb}{\mbox{$\times 10^{-2}$}}
\newcommand{\ddc}{\mbox{$\times 10^{-3}$}}
\newcommand{\ddd}{\mbox{$\times 10^{-4}$}}
\newcommand{\dde}{\mbox{$\times 10^{-5}$}}
\newcommand{\ddf}{\mbox{$\times 10^{-6}$}}

\begin{table}[bth]
\caption{Results of the  fits with fixed
 parameter    $\Delta =-0.76$ to the charged particle multiplicity
 distributions  measured by TASSO~[8] as a function of the centre of
mass energy $\sqrt{s}$ in restricted rapidity intervals
 $|y|<y_{cut}$. The last column gives the  $\chi ^2/NDF$ for the negative
 binomial fits taken from [8].}     
\begin{center}
\begin{tabular}{|c|c|c|c|c|c|c|}
\hline\hline             
$\sqrt s$ (GeV) &$y_{cut}$ & $k$ &$\epsilon_1$&$\epsilon_2$ &\axi & \axi (NB) \\
\hline\hline
  & 0.5 & 3 &0.151$\pm 0.022$ & 0 & 12.4/14
                                                   &15/13\\
  &    & 2 &0.391$\pm 0.017$ & 0.106$\pm 0.029$ &9.7/13 & \\ \cline{2-7}
14.0   & 1.0  & 5 &0.108$\pm 0.023$ & 0& 37.2/20   &27/19\\ \cline{2-7}
 & 1.5 & 7 & 0.164$\pm 0.023$ &0 &36.0/23  &33/22  \\
 &     & 5 & 0.372$\pm 0.045$ &0.170$\pm 0.061$& 36.8/21 & \\ \cline{2-7}
 & 2.0 & 9 &0.386$\pm 0.023$  &0 &85.0/25 &75/24 \\
 &     & 5 &0.239$\pm 0.021$  &0.544$\pm 0.018$&72.8/24& \\
\hline
 & 0.5 & 3 &0.181$\pm 0.022$  &0 & 3.0/14 & 4/13 \\
 &     & 2 &0.411$\pm 0.020$  &0.119 $\pm0.030$ & 4.8/13 &  \\ \cline{2-7}
22.0 & 1.0 & 3  &0.421$\pm 0.027$ &0&4.6/21 & 19/20 \\
     &     & 2  &0.470$\pm 0.045$ &0.379$\pm 0.043$ & 11.7/20 & \\ \cline{2-7}
 & 1.5 & 5  &0.283$\pm 0.028$ &0&6.5/26 & 8/25 \\
 &     & 5  &0.337$\pm 0.030$ &0.054$\pm 0.91$& 6.4/25 & \\ \cline{2-7}
 & 2.0 & 9  &0.186$\pm 0.025$ &0&14.5/28 & 13/27 \\
 &     & 8  &0.269$\pm 0.057$ &0.190$\pm 0.076$& 11.9/27& \\ 
\hline
 & 0.5 & 3 &0.153$\pm 0.011$ &0& 10.1/15    &22/14 \\ 
 &     & 2 &0.404$\pm 0.008$ & 0.109$\pm 0.015$& 6.9/14 & \\ \cline{2-7}    
34.8  & 1.0 &3  &0.387$\pm 0.013$  &0& 49.3/26  &102/25 \\
 &    & 2  & 0.462$\pm 0.018$& 0.403$\pm 0.017$ & 50.3/25 & \\ \cline{2-7}
 & 1.5 & 4&0.415$\pm 0.015$ &0&34.7/31 & 74/30 \\
 &     & 3&0.406$\pm 0.034$ &0.413$\pm 0.029$ & 30.7/29& \\ \cline{2-7}
 & 2.0 & 8&0.082$\pm 0.015$ &0&28.1/34 & 27/33 \\
 &     & 6&0.308$\pm 0.039$ &0.204$\pm 0.048$& 30.7/33 & \\
\hline
 & 0.5 & 3 &0.080$\pm 0.022$ &0& 6.7/17   &8/16\\
 &     & 2 &0.377$\pm 0.015$ &0.077$\pm 0.034$ & 7.1/16& \\ \cline{2-7}
43.6 & 1.0 &3  & 0.315$\pm 0.026$ &0& 25.8/27
                                                   &44/26 \\ 
 &     & 2 &0.451$\pm 0.035$ &0.399$\pm 0.034$& 21.4/26& \\ \cline{2-7}
 & 1.5 & 5 &0.105$\pm 0.027$ &0& 45.6/32  &39/31 \\ \cline{2-7}
 & 2.0 & 7 &0.117$\pm 0.027$ &0& 43.6/34
                                                   &39/33\\
\hline
\end{tabular}
\end{center}
\end{table}
\clearpage
\begin{table}[bth]
\caption{Results of the  fits with fixed parameter  $\Delta = -0.76$ to the 
          charged particle multiplicity distributions 
         measured by DELPHI~[9] and ALEPH~[11] 
 at $\sqrt{s} = 91.2$ GeV in the
 symmetric rapidity intervals \mbox{$|y|<y_{cut}$}.}

\begin{center}
\begin{tabular}{|c|c|c|c|c|c|}
\hline\hline             
Collaboration&$y_{cut}$&$k$ & $\epsilon_1$ &$\epsilon_2$& \axi      \\
\hline\hline
  & 0.5 & 2 &0.155$\pm 0.018$ &0& 55.8/21 \\ \cline{2-6}
DELPHI  & 1.0  & 2 &0.631$\pm 0.016$ &0&111.3/35   \\ \cline{2-6}
 & 1.5 & 3 & 0.481$\pm 0.014$ &0&279.4/44   \\
 &     & 2 & 0.362$\pm 0.019$ &0.730$\pm 0.018$& 295.6/43 \\ \cline{2-6} 
 & 2.0 & 4 &0.425$\pm 0.014$  &0&222.5/48   \\
 &     & 3 &0.346$\pm 0.030$ &0.606$\pm 0.027$& 218.9/47 \\ 
\hline
 & 0.5 & 2 &0.305$\pm 0.057$  &0& 5.1/19  \\ \cline{2-6}
ALEPH & 1.0 & 2  &0.753$\pm 0.068$ &0&11.6/33  \\
     &      & 2  &0.552$\pm 0.145$ &0.301$\pm 0.162$&10.8/32 \\ \cline{2-6}
 & 1.5 & 3  &0.519$\pm 0.056$ &0&32.7/39  \\
 &     & 2  &0.418$\pm 0.089$ &0.696$\pm 0.082$& 16.7/38 \\ \cline{2-6}
 & 2.0 & 3  &0.929$\pm 0.054$ &0&10.5/45  \\
 &     & 2  &0.335$\pm 0.066$ &0.981$\pm 0.057$& 37.0/44 \\
\hline
\end{tabular}
\end{center}
\end{table}
\begin{table}[bth]
\caption{Results of the  fits by the superposition of the two weighted
 MNBDs
   to the charged particle multiplicity
 distributions  measured by DELPHI~[9] 
 at $\sqrt{s} = 91.2$ GeV in restricted rapidity intervals
 $|y|<y_{cut}$.}              

\begin{center}
\begin{tabular}{|c|c|c|c|c|c|c|}
\hline\hline             
 $y_{cut}$ & $k$ & $\alpha$ &$\epsilon_1$&$r_1$ &$r_2$ & \axi  \\
\hline\hline
0.5 &5&0.735$\pm 0.022$&0.058$\pm 0.034$& 6.87$\pm 4.62$ &15.28$\pm 9.36$&
 11.0/18\\
1.0 &5&0.727$\pm 0.013$&0.140$\pm 0.024$& 6.08$\pm 1.27$ &12.31$\pm 2.29$&
 27.9/32\\
1.5 &8&0.681$\pm 0.010$&0.062$\pm 0.022$& 6.50$\pm 1.24$ &13.03$\pm 2.22$&
 51.1/41\\
2.0 &9&0.570$\pm 0.011$&0.131$\pm 0.024$& 7.05$\pm 1.47$ &13.19$\pm24.92$&
 46.7/45\\
\hline
\end{tabular}
\end{center}
\end{table}
\begin{table}[bth]
\caption{Results of the  fits by the superposition of the two weighted
 MNBDs
   to the charged particle multiplicity
 distributions  measured by DELPHI~[9] at $\sqrt{s} =91.2$ GeV
in restricted rapidity intervals
 $0<y<y_{cut}$.}
                  
\begin{center}
\begin{tabular}{|c|c|c|c|c|c|c|}
\hline\hline             
 $y_{cut}$ & $k$ & $\alpha$ &$\epsilon_1$&$r_1$ &$r_2$ & \axi  \\
\hline\hline
0.5 &4&0.790$\pm 0.025$&0.060$\pm 0.039$& 3.96$\pm 2.24$ &9.58$\pm 6.78$&
 3.6/12\\
1.0 &4&0.848$\pm 0.009$&0.061$\pm 0.029$& 9.50$\pm 4.52$ &23.83$\pm 11.38$&
 35.4/25\\
1.5 &6&0.769$\pm 0.008$&0.065$\pm 0.022$& 8.05$\pm 2.98$ &20.51$\pm 7.03$&
 55.8/29\\
2.0 &9&0.685$\pm 0.008$&0.055$\pm 0.021$& 7.84$\pm 3.27$ &19.51$\pm 7.48$&
 75.5/30\\
3.0 &9&0.423$\pm 0.017$&0.219$\pm 0.027$& 2.24$\pm 0.41$ &4.51 $\pm 0.62$&
 36.5/30\\
5.0 &9&0.563$\pm 0.061$&0.467$\pm 0.025$& 1.44$\pm 0.19$ &2.01 $\pm 0.16$&
 31.5/30\\
\hline
\end{tabular}
\end{center}
\end{table}
\begin{table}[bth]
\caption{Results of the  fits with fixed
 parameter 
         $\Delta =-0.76$ to the 
          charged particle multiplicity distributions 
         in a single hemisphere for $e^+e^-$ annihilation into hadrons.}       
\begin{center}
\begin{tabular}{|c|c|c|c|c|c|}
\hline\hline             
Experiment&$\sqrt s$ (GeV)&$k$ &$\epsilon_1$ &$\epsilon_2$& \axi      \\
\hline\hline
  & 14.0 & 6 &0.566$\pm 0.009$ &0& 10.5/17 \\
  &      & 4 &0.568$\pm 0.027$ &0.292$\pm 0.024$ & 7.1/16 \\ \cline{2-6}
TASSO~[8]   & 22.0  & 8 &0.456$\pm 0.010$ &0&5.5/17   \\
 &       & 6 &0.535$\pm 0.024$ & 0.162$\pm 0.030$& 5.0/16 \\ \cline{2-6}
 & 34.8 & 8 & 0.470$\pm 0.005$ &0&26.4/23   \\
 &      & 6 & 0.541$\pm 0.015$ &0.196$\pm 0.017$ & 24.7/22 \\ \cline{2-6}
 & 43.6 & 9 & 0.376$\pm 0.010$  &0&10.6/23  \\ 
 &      & 6 & 0.483$\pm 0.033$  &0.253$\pm 0.033$& 12.5/22 \\ 
\hline
HRS~[7] & 29.0 & 9 &0.487$\pm 0.011$  &0& 17.8/18  \\
 &  &            5 &0.576$\pm 0.047$  &0.321$\pm 0.039$ & 24.5/17 \\
\hline
OPAL~[10] with $K_s$, $\Lambda$ & 91.2 & 5  &0.653$\pm 0.018$ &0&18.7/34  \\
 &   &           5 &0.667$\pm 0.080$ & 0.245$\pm 0.078$ & 12.1/33 \\           
\hline
OPAL~[10] no $K_s$, $\Lambda$ & 91.2 & 5  &0.782$\pm 0.018$ &0&7.6/32  \\
 &   &           5 &0.684$\pm 0.086$& 0.150 $\pm 0.100$& 7.0/31 \\
\hline
DELPHI~[9] & 91.2 & 6  &0.671$\pm 0.008$ &0& 98.7/34  \\
 &   &           4 &0.309$\pm 0.023$ & 0.729$\pm 0.018$& 80.7/33 \\   
\hline
  &   & 6 &0.844$\pm 0.011$ &0& 13.3/27  \\
DELPHI~[12], $b\overline{b}$-events& 91.2 &5 &0.592$\pm 0.068$&0.452$\pm 0.062$
 & 7.9/26 \\
\hline
 & & 6 & 0.691$\pm 0.012$ &0& 24.9/29 \\ 
DELPHI~[12], all flavours&91.2&5&0.506$\pm 0.057$&0.412$\pm 0.048$&
 20.1/28   \\
\hline
\end{tabular}
\end{center}
\end{table}
\begin{table}[bth]
\caption{Results of the  fits with fixed
 parameter 
         \mbox{$\Delta =-0.76$} to the 
          charged particle multiplicity distributions 
         measured by H1~[13] in restricted pseudorapidity intervals
 $1<\eta <\eta_{cut}$.}
\begin{center}
\begin{tabular}{|c|c|c|c|c|c|}
\hline\hline             
 W(GeV)&$\eta$ interval& $k$ & $\epsilon_1$&$\epsilon_2$ & \axi      \\
\hline\hline
  & $1\div 2$ & 3 &0.038$\pm 0.073$ &0& 0.8/13 \\
  &           & 1 &0.547$\pm 0.015$ &0.303$\pm 0.047$& 14.3/12 \\ \cline{2-6}
$80\div 115$  & $1\div 3$ & 4 &0.100$\pm 0.076$ &0&1.8/17   \\
  &           & 3 &0.339$\pm 0.102$ & 0.167 $\pm 0.134$& 2.2/16 \\ \cline{2-6}
 & $1\div 4$ & 9 & 0.026$\pm 0.079$ &0&0.5/18   \\
  &           & 5 &0.317$\pm 0.215$ &0.229$\pm 0.228$& 0.7/17 \\ \cline{2-6}
 & $1\div 5$ & 9 &0.162$\pm 0.044$  &0&1.6/18   \\
  &           & 3 & 0.346$\pm 0.056$ & 0.627$\pm 0.060$ & 9.9/17 \\ 
\hline
 & $1\div 2$ & 2 &0.199$\pm 0.035$  &0& 6.9/14  \\ 
 &           & 1 &0.533$\pm 0.026$  &0.285$\pm 0.027$&62.5/13 \\ \cline{2-6}
$115\div 150$ & $1\div 3$ & 3  &0.178$\pm 0.090$ &0&2.8/18  \\
 &           & 1 &0.588$\pm 0.060$  &0.385$\pm 0.055$&113.3/17 \\ \cline{2-6}
 & $1\div 4$ & 7  &0.043$\pm 0.056$ &0&4.5/20  \\
 &           & 7  &0.174$\pm 0.048$ &0.122$\pm 0.110$ & 3.4/19 \\ \cline{2-6}
 & $1\div 5$ & 9  &0.154$\pm 0.047$ &0&7.6/21  \\
 &           & 4  &0.319$\pm 0.058$ &0.515$\pm 0.056$& 12.8/20 \\ 
\hline
 & $1\div 2$ & 3  &0.028$\pm 0.092$ &0&2.9/14  \\
 &           & 3  &0.182$\pm 0.055$ &0.066$\pm 0.042$& 1.0/13 \\ \cline{2-6}
$150\div 185$& $1\div 3$&3 &0.195$\pm 0.116$ &0& 4.0/20  \\
 &           & 2  &0.442$\pm 0.170$ &0.322$\pm 0.097$& 3.0/21 \\ \cline{2-6}
 & $1\div 4$ & 6  &0.064$\pm 0.083$ &0&4.4/22  \\
 &           & 6  &0.162$\pm 0.083$ &0.178$\pm 0.097$& 3.0/21 \\ \cline{2-6}
 & $1\div 5$ & 9  &0.088$\pm 0.064$ &0&3.4/22 \\
 &           & 9  &0.218$\pm 0.034$ &0.154$\pm 0.088$&1.5/21 \\
\hline
 & $1\div 2$ & 3  &0.034$\pm 0.096$ &0&2.4/14 \\
 &           & 3  &0.173$\pm 0.064$ &0.056$\pm 0.041$& 0.6/13 \\ \cline{2-6}
 $185\div 220$ & $1\div 3$& 3 &0.188$\pm 0.096$ &0& 3.8/21 \\
 &           & 2  &0.366$\pm 0.075$ &0.322$\pm 0.059$& 8.4/20 \\ \cline{2-6}
 & $1\div 4$ & 5  &0.126$\pm 0.091$ &0&9.2/22 \\
 &           & 4  &0.232$\pm 0.060$ &0.300$\pm 0.067$& 8.1/21 \\ \cline{2-6} 
 & $1\div 5$ & 9  &0.169$\pm 0.074$ &0&8.7/23 \\
 &           & 5  &0.233$\pm 0.060$ &0.502$\pm 0.064$ &6.2/22 \\
\hline
\end{tabular}
\end{center}
\end{table}
%
%
\end{document}